# Dynamic State Estimation for Power System Control and Protection

*IEEE Task Force on Power System Dynamic State and Parameter Estimation*
Yu Liu, *Member, IEEE*, Abhinav Kumar Singh, *Member, IEEE*, Junbo Zhao (Chair), *Senior Member, IEEE*, A. P. Meliopoulos, *Fellow, IEEE*, Bikash Pal, *Fellow, IEEE*, M. A. M. Ariff, *Member, IEEE*, Thierry Van Cutsem, *Fellow, IEEE*, Mevludin Glavic, *Senior Member, IEEE*, Zhenyu Huang, *Fellow, IEEE*, Innocent Kamwa, *Fellow, IEEE*, Lamine Mili, *Life Fellow, IEEE*, Saleem Mir, *Member, IEEE*, Ahmad Taha, *Member, IEEE*, Vladimir Terzija, *Fellow, IEEE*, Shenglong Yu, *Member, IEEE*

*Abstract*—Dynamic state estimation (DSE) accurately tracks the dynamics of a power system and provides the evolution of the system state in real-time. This paper focuses on the control and protection applications of DSE, comprehensively presenting different facets of control and protection challenges arising in modern power systems. It is demonstrated how these challenges are effectively addressed with DSE-enabled solutions. As precursors to these solutions, reformulation of DSE considering both synchrophasor and sampled value measurements and comprehensive comparisons of DSE and observers have been presented. The usefulness and necessity of DSE based solutions in ensuring system stability, reliable protection and security, and resilience by revamping of control and protection methods are shown through examples, practical applications, and suggestions for further development.

## I. Introduction

Today's power systems are witnessing a rapid transition in generation technology from coal and gas-based non-renewable generation to wind and solar energy based renewable generation. Energy storage and modern power electronic loads penetrate power systems rapidly as well. As more and more of such converter-based resources (CBRs) are connected to the network, two major challenges emerge: (a) system response is faster and, therefore, its control should have a commensurate response speed, and (b) legacy protection functions that rely on the characteristics of conventional power systems (high fault currents and fault characteristics associated with synchronous generation) can be inadequate. Solutions to these challenges can be sought in dynamic state estimation (DSE) applications [1]. Specifically, DSE provides real-time operating states of the system at fast rates [2], which can be utilized to fulfill the requirements for protection and control of modern power systems.

Traditional schemes for power system control and protection are primarily based on a deterministic system model with the majority of electricity coming from a few centralized and synchronous sources of generation [3-6]. Such a system model has gradually become out of context due to the distributed, unpredictable, and intermittent nature of renewable energy sources. In addition, stability, control and protection challenges introduced by the increased renewable integration include: 1) reduced system inertia, 2) limited observability/controllability of system dynamics and 3) extreme variation in the timescales of system dynamics – from a few milliseconds or lower in the case of CBRs to a few minutes or higher in the case of boiler and long-term dynamics of synchronous generation. To deal with these challenges, it is necessary to consider the control of each dynamic component in the system individually, and how different components influence each other's dynamics, and how they should be controlled together holistically. As DSE provides state estimates at the required timescales, it can serve as a versatile tool to holistically control system trajectory, ensuring rotor-angle stability, frequency stability, voltage stability, and benefiting controls of CBRs.

With the increasing penetration of CBRs in power systems, traditional schemes for power system protection face the following challenges: 1) Many traditional protection schemes depend on abrupt changes of voltages/currents during faults, e.g. overcurrent/undervoltage relays to detect faults; however, these characteristics may not be valid in systems with high penetration of CBRs [3]; 2) Phasor domain quantities are usually utilized in traditional relays, which can cause misoperation of relays during complex and unusual system transients in CBR-dominated power systems; 3) Traditional protection schemes usually require complex coordination among relays, e.g. time overcurrent relays and 3-step distance relays, resulting in risks of mis-coordination; 4) Hidden failures, such as failures of instrumentation channels, can lead to misoperation of protection relays [4]. DSE is a valuable tool to overcome the above limitations. First, DSE can accurately track complex dynamics and provide accurate estimates of the internal states of the system, enabling precise extraction of fault characteristics in both the phasor domain and time domain [1]. In addition, for a specific protection zone, DSE can formulate the protection logic by systematically checking the consistency between the measurement and the dynamic model of the protection zone without coordination among relays [7]. Finally, with redundant measurements, DSE is able to identify and reject bad data, and therefore prevent relays from misoperation during hidden failures.

With the development of advanced measurement devices and substation automation, high-quality synchrophasor measurements and synchronized sampled value (SV) measurements provide more information at higher rates and enable DSE-based advanced control and protection schemes. Towards this end, this paper summarizes the joint efforts of the Task Force on Power System Dynamic State and Parameter Estimation, with an emphasis on DSE for power system control and protection. It has the following new insights: 1) the original DSE formulation for electromechanical dynamics has been extended to consider electromagnetic dynamics, such as those from CBRs (Section II); 2) the relationships and differences among DSE and observers have been compared and discussed (Section III); 3) the roles of DSE for control and protections have been extensively discussed with the support of numerical results (Sections IV and V, respectively); and 4) future research directions have been discussed to pave the way for further development (Section VI).

## II. DSE Formulation: Sampled Value Measurements versus PMU Measurements

Traditionally, power systems are dominated by synchronous generators. For these power systems, electromechanical oscillations with periods of a few seconds are very important. Although the detailed synchronous machine dynamic models involve electromagnetic transients, they are too fast as compared to the electromechanical oscillations and thus are neglected in the traditional dynamic studies and stability assessment [5]. With the increasing penetration of CBRs, such as distributed energy resources (DERs) and FACTS devices, to cite a few, the system dynamic responses are heavily dependent on the fast-response power electronic devices and their controls, and converter-induced dynamics and stability issues start to dominate the system [5][8][9]. In addition, power electronics devices cause waveform distortions and deviations from near sinusoidal waveforms. Note that the time-scale of the CBRs can range from a few microseconds to several milliseconds, due to switching operations of the power electronics. Conventional phasor representation or quasi sinusoidal approximation is usually used for the study of electromechanical oscillations accounting for the synchronous machine oscillations and converter control dynamics (the dynamic phasors [10] or average models [11] could simplify the design of protection and control strategies for CBR systems). However, conventional phasor representation may not be suitable for the study of electromagnetic phenomena which represent the fast dynamics of the system. Hence, the formulation, measurement requirements, and potential applications of DSE are revisited in this paper, and the requirements for each application are defined.

Irrespective of the different time scales of power system dynamics, they can be described by differential and algebraic equations (DAEs). Note that for models described through partial differential equations (PDEs), such as wave propagations in transmission lines, appropriate discretization methods need to be utilized to convert PDEs into (1).

$$\begin{cases} \dot{x}(t) = f(x(t), y(t), u(t), p(t)) \\ 0 = g(x(t), y(t), u(t), p(t)) \end{cases} \quad (1)$$

where $x$ is the state vector; $y$ is the algebraic variable vector; $u$ is the input vector, $p$ is the parameter vector; and $f$ and $g$ are nonlinear vector-valued functions. For power system applications, the measurements are sampled in a discrete manner and thus, (1) needs to be discretized to be compatible with online measurements. To this end, the measurement function at time instant $k$ can be written as $z_k = h(x_k, y_k, u_k, p_k)$, where $z_k$ can come from PMUs, merging units (MUs), digital fault recorders, and in general intelligent electronic devices (IEDs); and $h$ is a nonlinear vector-valued function.

For capturing the electromechanical transients, phasor representation is leveraged while fast-electromagnetic transients are neglected. Measurement vector $z_k$ usually includes time-synchronized phasor measurements from PMUs, such as voltage and current phasors, the calculated real and reactive powers, frequency and rate of change of frequency (RoCoF); and the state vector $x$ includes internal dynamic variables of synchronous machines and dynamic loads. Note that the phasor measurements provide the fundamental frequency phasor voltages and currents, while the electromagnetic transients are filtered out. Thus, the model in (1) represents the electrical quantities with phasors, and the electromechanical system and control dynamics are formulated with differential equations.

To capture fast electromagnetic transients in a power electronics-dominated power system, there is an increasing need to use SV measurements directly. The synchronized SV measurements contain rich time-domain information and can be obtained from MUs, which can be standalone devices or embedded in non-conventional instrument transformers and other apparatus. The standard SV sampling rates in MUs are 80 or 256 samples per cycle according to IEC61850-9-2LE standard [6]. Note that phasors and harmonics can be computed from SVs. Furthermore, electromagnetic transient models must be adopted to represent the fast dynamics of system components. The latter can include any components with electromagnetic transients, such as generators, transmission lines, transformers, CBRs, among others [7]. In this case, the voltages and currents are expressed using instantaneous SVs; the electromagnetic transients, such as fast electromagnetic transients of voltages or currents, are considered. Measurement vector $z$ includes SV measurements, and the state vector $x$ consists of instantaneous voltages, currents, generator speeds, and other internal states of the dynamic components.

The applications of DSE using PMU measurements for electromechanical transients include both control and protection. For control applications, DSE contributes to the enhancement of the observability of the system dynamics and the validation and calibration of the control models. Furthermore, it provides essential feedback state signals and accurate measurements for controls. It is shown in the literature that DSE-based out-of-step protection provides more benefits for fast and reliable relay actions than traditional approaches.

DSE using SV measurements for both electromagnetic and electromechanical transients also enables advanced control and protection applications. For the control applications, it can 1) provide essential state feedback signals for CBR control, including both grid following and grid forming controls to enhance system stability; 2) identify unknown parameters and help calibrate electromagnetic models; 3) detect and diagnose anomalies locally, especially for converter-interfaced resources; and 4) provide accurate frequency measurements even in the presence of large disturbances. The widely used phase-lock-loop is well-known to be vulnerable to large disturbance and can trigger erroneous relay actions, for example, see the 2016 California blue-cut fire events [8]. For protection applications, it allows us to 1) design protective relays with improved dependability, security, sensitivity, and speed; 2) develop fault locators that work with short data window with improved fault location accuracy; and 3) enable cyber intrusion/hidden failure detection of protective relays. These issues will be discussed in subsequent sections.

## III. Controller Options: DSE versus Observer

Both observers and DSE can be used to provide state feedback signals for control. In this section, the advantages and disadvantages of these two methods are discussed.

**Operational principle**: Observers are based on sensor outputs measured from the physical system and are also usually based on the Luenberger criterion [12]. An observer can also be designed using an overdetermined Koopman-based model of the process [13]. DSE is based on the minimum variance estimation criterion from a statistics point of view and it fuses both the physical model predictions and the sensor outputs. Although these two approaches fundamentally differ in their assumptions and algorithmic details, both can be used to estimate the state of a dynamic system.

**Addressing stochastic systems**: Luenberger-type observers are typically restricted to the deterministic case and no statistics of the model and measurement are involved. Most observer designs assume that the unknown signals (process and measurement noise, unknown inputs, etc.) are bounded. Besides, gain matrix design for an observer can be challenging for large nonlinear systems due to high nonlinearity in state transition[14]. A good observer design can provide robustness to exogenous disturbances, though its performance cannot adapt to time-varying changes unless it evolves with time. Furthermore, if the noise distribution deviates from assumptions such as Gaussian distribution, an observer may obtain better performances than the traditional Kalman filter based DSEs since an observer only requires the bounds for noise. Instead of Gaussian assumption based DSE, if robust DSE is adopted [15], it typically provides better estimates than an observer subject to stochastic noise. DSE considers both process and measurement noise and derives the optimal estimation criterion to minimize estimation variance, thereby effectively dealing with system stochasticity.

**Sensitivity to outliers**: Measurements are frequently subjected to outliers and the model outputs can be corrupted by gross errors due to control failures, incorrect model inputs, parameter errors, etc. [15]. In the presence of outliers, both state observer and DSE may provide biased results. To mitigate the influence of outliers, an observer needs to increase the assumed error bounds, which would significantly decrease its performance in the absence of outliers. In contrast, by adopting a statistical test or developing robust DSEs, outliers can be automatically detected and suppressed without affecting DSE performance when there are no outliers [1] [15].

**Computational efficiency**: The majority of DSE designs recursively compute gains; this involves matrix multiplications and inversions at each time-step. In contrast, once the observer is designed offline, state estimation can be performed with a significantly smaller number of matrix multiplications, as the calculations of observer gain are usually done offline assuming certain error bounds of process and measurement models [16]. This offline calculation involves computationally costly linear matrix inequalities and convex semidefinite programs. Thus, recalculation of observer gains is time-consuming in the presence of changes in the network topology or parameters, whereas DSE can update such new information more efficiently at each time-step.

**Sensor requirements**: Both DSE and observers require the system/states to be observable as a prerequisite from the dynamical system perspective [1]. For DSE, more sensors would lead to better statistical efficiency of the state estimates. Observers usually suffer from a key limitation regarding the required number of sensors, which can lead to infeasible observers [17]-[18], theoretically infinite estimation error, or practically unusable estimates. Kalman filters, however, still produce some useful results even with a limited number of sensors.

**Handling system nonlinearity**: Both DSE and observers can be designed to deal with nonlinear systems. However, designing a good observer when the system is subjected to complex nonlinearities is very challenging as it involves nonlinear optimization [18]. Also, the Jacobian matrix is needed for some nonlinear observer designs. However, magnetic saturation in synchronous machines is frequently encountered in power systems [2], making infeasible the computation of Jacobians and subsequent observer design. In contrast, there exist derivative-free nonlinear Kalman filter-based DSEs, such as unscented Kalman filter, ensemble Kalman filter, particle filter, etc. [1], which avoid the use of Jacobian matrix and thus handle system nonlinearity better.

**Sensitivity to initial conditions**: For DSE, a good initial condition is usually needed, otherwise, it takes some time to converge to the actual value. This, in general, is not an issue since DSE runs continuously and in the long run, once it has converged, the initial condition is the state estimate of the system at the prior time step, which is an accurate initial condition. For state observers, irrespective of the initial condition, fast convergence of the estimated states to their accurate values is guaranteed as long as a proper observer gain is determined, and the system is observable. It should be noted that practically, a reasonable initial condition can typically be obtained by utilizing power flow calculations or state estimation, or engineering judgments or experiences.

## IV. Control Applications of DSE

**Overview of control**: Control of different components in a power system is performed by controlling the devices which govern their power and voltages. For a synchronous generator, these devices are the governor (which controls the mechanical power input) and the automatic voltage regulator (AVR) (which controls the excitation voltage). For CBR, on the other hand, these devices are the power electronics converters, which have a few control options, for example, maximum power tracking and voltage control, real and reactive power control, real power and voltage control, and AC/DC voltage control. In addition, depending on operating conditions, the controls can switch to low voltage ride through logic, storage control, etc. Since synchronous generator controls are slower than converter controls, DSE must provide feedback to the controllers on a commensurate time scale. DSE operating at a cycle scale can meet the requirements of controllers for synchronous machines and converter control. For other applications such as protection and associated controls, DSE needs to operate at faster time scales and typically uses SV measurements or dynamic phasors. An overall scheme for implementation of DSE-based control of power systems is shown in Fig. 1.

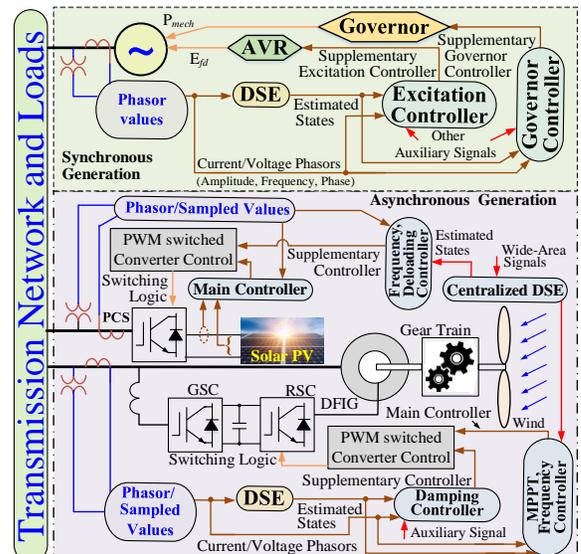

Fig. 1.  Example implementation of DSE-based control architecture

### A. DSE-Based Control using PMU Measurements

#### 1) DSE-based Control of Rotor Angle Stability

As synchronous generators form the backbone of today's AC power system, they are the main points of control-actuation and can be directly controlled using excitation systems & governors. Other important actuators are the transmission paths, controlled using FACTS devices and transformer taps. DSE-based control design can be centralized, decentralized, or hierarchical (Fig. 2).

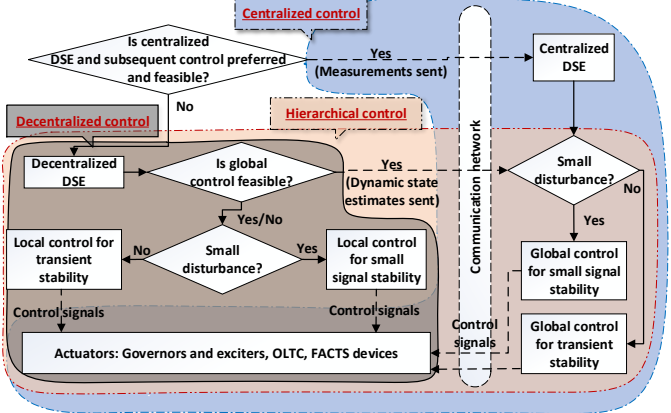

Fig. 2. A flowchart of DSE-based control decision and design process for rotor angle stability

*Centralized and Hierarchical Control*

In DSE-based centralized control, measurements from across the system are transmitted to a central location, and DSE for the whole system is performed [19]-[21], as shown in Fig. 2. A control law is then obtained using the system model and the dynamic state estimates via various control techniques, including $L_\infty$ robust control that models disturbances as $L_\infty$ bounded inputs and finds a state-feedback control law using nonconvex optimization [19]; linear quadratic regulator (LQR) where the quadratic costs of control and state-deviation are optimized with pole-placement for crucial system eigenvalues using state-feedback [20]; and residue-based control, where pole-placement is explicitly done using power system stabilizers (PSSs)/power oscillation dampers [21]. Some limitations of centralized control that restrict its field implementation are: communication latencies can impact performance, an accurate model of the whole system at the central location is required, and communication failures and bottlenecks can create serious issues.

A partial solution to the problems of centralized control is hierarchical control in which decentralized DSE is performed at machine locations and the estimates are sent to a central location (or PMU measurements are sent to a central location for performing DSE in a decentralized manner using a federation of estimators [22]). Global control laws are then developed at the central location along with local control laws for decentralized locations. The local and global controls form the two levels of hierarchical control, as shown in Fig. 2, and are implemented using actuators, such as FACTS devices and excitation systems of synchronous generators [23]-[24]. In the worst-case scenario, such control can also be used to shed run-away generators on the fly to prevent instability [25]. The hierarchical control approach also requires knowledge of a complete system model at the central location.

*Decentralized Control*

In decentralized control, each generator is controlled independently, requiring only local measurements for both DSE and control. By controlling local machine dynamics, global system dynamics are controlled. This eliminates communication requirements, but in many cases, it suffers from limited system-level controllability/observability. As a result, special techniques are needed to establish system stability. Both linear and nonlinear methods can be used to implement such control.

Linear methods of DSE based decentralized control are valid only for small-signal dynamics [5], and are similar to the traditional PSSs: in PSSs the control parameters are tuned offline, while DSE-based methods do so online. Such a 'dynamic tuning' of control parameters can be done using various methods, such as decentralized $L_\infty$ [19], in which a convex approximation of centralized $L_\infty$ formulation ensures that only local machine states are needed for control; and extended LQR [26], in which the costs of voltage phasors (as exogenous inputs) are included in the LQR costs to find control gains. A drawback of linear methods is that the asymptotic stability of the whole system cannot be guaranteed. However, this is usually not an issue as linear methods are applicable only for small disturbances, which usually do not alter the asymptotic stability.

Another linear control approach is to use data-driven DSE, where local measurements are used to find a linear Koopman predictor [13] to mimic system dynamics, which is then used to perform DSE (similar to an observer) and to design model predictive control (MPC). The estimated rotor speed of each generator is used as a feedback signal in the MPC for excitation control. Fig. 3 shows that the proposed Koopman MPC (KMPC) provides a much better oscillation damping performance as compared to the classical PSS, after a 100 ms symmetrical three-phase fault at the infinite bus in a single machine infinite bus system. This example uses a highly accurate Koopman model trained with random excitations. Its dimension is N=108 (8 physical states plus 100 Radial Basis Functions) [13].

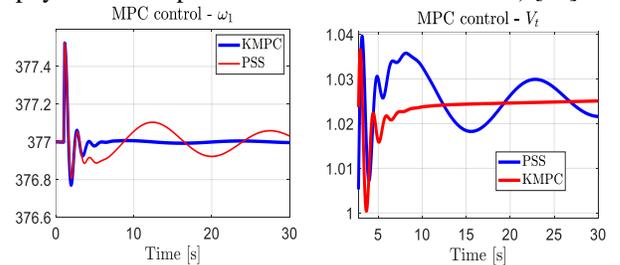

Fig. 3. Rotor speed response of KMPC vs PSS

Nonlinear control methods are needed to ensure the stability of the system against large disturbances, i.e. when a linear approximation of the system is no longer valid. These methods either find a partially linear transformation of system dynamics, called feedback linearization [27]-[28], or find a positive scalar with a negative time derivative, called a Lyapunov function [29]-[30]. Feedback linearization is easier to formulate and use than Lyapunov functions, but the latter has better asymptotic stability characteristics. In [27], detailed generator modeling has been used to implement feedback linearization based excitation control to ensure transient stability, while [28] considers detailed load modeling for the same. In [29], a Lyapunov function is constructed for excitation and governor control, and [30] proposes an optimal Lyapunov formulation for excitation control and uses neural networks for computational efficiency.

Although nonlinear control works well even for small disturbances, its control costs (given by the time integral of a weighted sum of the squares of control inputs) are much higher than the linear control costs. Hence, it is practical to activate nonlinear control only for large disturbances. Based on this idea, a hybrid decentralized control is proposed in which RoCoF estimation is used to infer if a large disturbance has occurred. To do this, the DSE-based RoCoF estimate obtained using [31] is multiplied with the machine's inertia to get a switching signal: if the absolute value of this signal is greater than a predefined value (say 0.5 p.u.) for two cycles, then the control is switched to the nonlinear control given in [27]. Control is switched to the linear control given in [26] only if the switching signal stays within ±0.5 p.u. continuously for 5 s. The comparison of performances of the hybrid control and PSS control (system model, simulation conditions, and controller parameters are as those in [27]) shows that hybrid control can ensure both small-signal stability and transient stability of the system (Fig. 4).

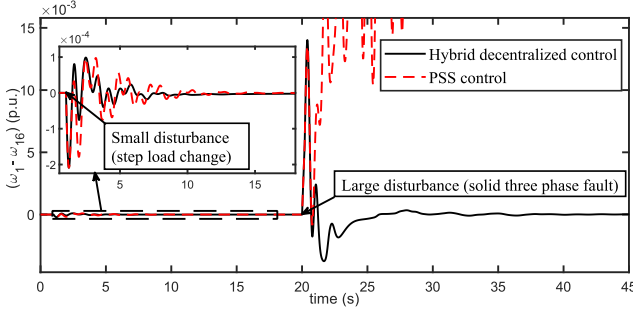

Fig. 4. Control performance of decentralized hybrid-control

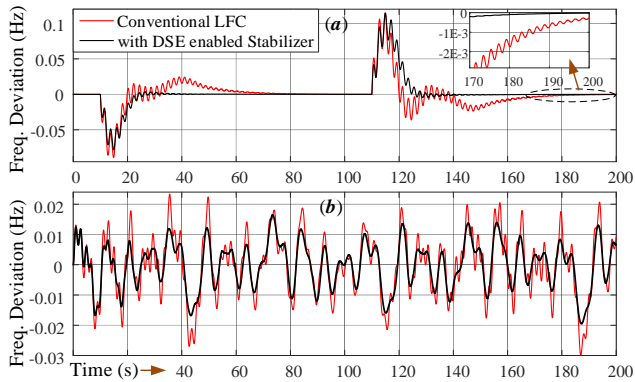

Fig. 5. Frequency Stability: DSE-Stabilizer vs conventional LFC

*2) DSE-based Control of Frequency Stability*

Unpredictable load fluctuations, intermittency due to stochastic generation, and random system disturbances may result in continuous fluctuations in system frequency and tie-line-power exchanges causing them to deviate from their nominal values. These variations are corrected by governor-based load frequency controller (LFC) action [32]. The performance of LFCs is significantly affected by the inherent governor deadband (GDB) and generation rate constraint (GRC) [33]-[34]. In contrast, DSE-enabled LFC [35] (Fig. 1) is not affected by the GDB or GRC and uses PMU measurements to derive the auxiliary signal for the governor loop. However, communication impediments hinder the applicability of such LFCs. To address this issue, it is proposed that traditional LFCs should be assisted by a supplementary stabilizer, in which turbine-governor dynamics are estimated using decentralized DSE to derive the control signal while load disturbances and stochastic generation are modeled as unmatched perturbations. This improves the frequency stability margins compared to observer-based LFCs (such as [36]) and is illustrated in Fig. 5, in which the IEEE 39 bus system is subjected to stepped load changes with GDB = 0.12% and GRC= 10% per minute (as in [33]) in scenario 1 and gusty wind (Van'der Hoven spectrum [34]) in scenario 2, and the resulting frequency deviations in area 1 have been plotted.

*3) DSE-based Control of Voltage Stability*

Real-time voltage instability detection and control methods that are the best candidates to take advantage of DSE are those that use snapshots of evolving system trajectories, captured either by dedicated measurement/communication system or by a tracking state estimator [37]-[39]. In principle, those methods can use the system snapshots provided by a DSE giving access, for instance, to the excitation system status of generators (namely: is the generator controlling its voltage or is it under field or armature current limit?) and system frequency. The same holds for other voltage controlling components such as static VAr compensators or STATCOMS. Long-term voltage instability detection and control can use originally proposed approaches (like sensitivities in [37] and modal analysis in [39]). In addition to the dynamic states, the network state and topology are required to monitor, detect, and control voltage instability. In this respect, the local DSE approaches discussed in [1] are combined with network state estimation, which can be fully based on PMUs or multi-rate measurements [38]. In principle, for long-term voltage instability, the system state can be updated at low rates (less than 10 times per second). One of the advantages of DSE is its ability to provide full state estimates, thus enabling full-state feedback MPC controllers [40].

An example of emergency control against long-term voltage instabilities [41], which can be adapted to take advantage of DSE outputs and be used for long-term instability detection, is illustrated in Fig. 6.

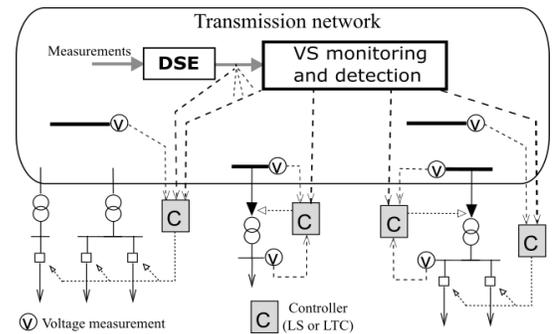

Fig. 6. A framework for two-level long-term voltage stability (VS) monitoring, instability detection, and emergency control

It is a two-level scheme. The upper level is in charge of wide-area monitoring, while the lower-level controllers provide remedial actions such as Load Shedding (LS) or modified Load Tap Changer (LTC) control. In an emergency voltage situation, the upper level relies on the voltage instability detection method of [37] (DSE can replace the dedicated PMU measurements proposed in[37]) and provides the lower-level controllers with minimum transmission voltages to maintain. Those voltages correspond to the point where long-term voltage instability is detected [41].

In terms of short-term voltage instability detection and control, the voltage control statuses of synchronous generators or compensators are crucial information. The phenomenon of concern is faster and a DSE used for this purpose must provide the system dynamic states at a high rate (10 to 60 times per second); a PMU-based network state estimation appears to be mandatory for this application.

Increased penetration of grid-connected CBRs (particularly when connected to weak grids) brings concerns to voltage instability (both long- and short-term). This requires even more rigorous metering infrastructure and DSE should use a detailed model with the full complement of state variables for the CBRs to solve voltage instability problems.

*4) DSE for CBR Control*

The intermittent and asynchronous nature of variable CBRs, mainly wind and solar, poses several pressing challenges in today's energy systems research [5][9], especially for CBR-DSE and control. Several methods for DSE based control of CBRs have been proposed in the literature. A UKF-based DSE is developed to estimate doubly-fed induction generator (DFIG) flux dynamics and this allows for the development of a flux estimation based control scheme [42], which achieves better fault recovery response than traditional control methods. In [43], DSE has been used to estimate the electromechanical dynamics of a DFIG and the estimated states have been used to derive the supplementary signal for damping of electromechanical oscillations (Fig. 1). DSE-based sliding mode control for DFIG-integrated power systems is developed in [44] for maximum energy extraction and power quality enhancement. The idea is further extended to the DSE-based frequency restoration method considering solar irradiance variations [45]. DSE based control can also be used to damp sub-synchronous oscillations in series-compensated lines with wind generation [46].

*B. DSE-based Control using SV Measurements*

Traditional generator controls are mostly decentralized as the local frequency and voltage information acts as a medium to bring the information of the rest of the grid to the local generator. Modern power systems are evolving with a lower inertia and more complex transients with CBRs. In this case, controls with local frequency and voltage information may not be sufficient; remote side information such as frequency, RoCoF, and waveform distortion can also help minimize the transients and prevent damage/shut-down of converters during disturbances. However, remote side information usually requires communication channels, resulting in increased cost and compromised reliability of the system. With the help of SV measurements and the accurate time-domain transmission line model, DSE can effectively estimate the voltages and currents at the remote side of the transmission line using local information only, without any physical communication channels between the two terminals of a transmission line [47]-[48]. Afterward, remote side information such as frequency, RoCoF, and waveform distortion is extracted from the estimated voltages and currents and is utilized as the input to CBR control. Fig. 7 depicts an example of a DSE-based converter control system using SV measurements.

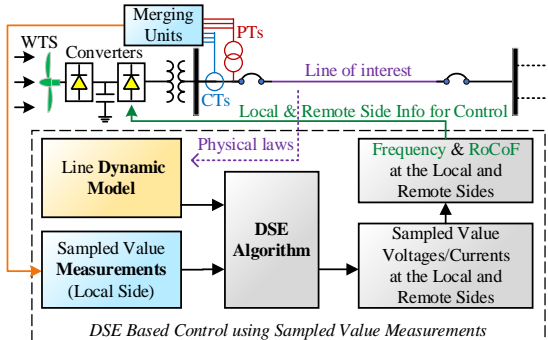

Fig. 7.　DSE based converter control system using SV measurements

In this example, the local voltage and current SV measurements are first obtained via potential transformers (PTs), current transformers (CTs), and MUs. Next, using the DAEs of the line of interest, DSE estimates the dynamic states of the system, including SV voltages and currents at the remote side. Finally, the frequency and the RoCoF at the local and remote sides are extracted from the estimated SV measurements. These are fed into the local converter controller to initiate supplementary control using local and remote frequency and RoCoF to minimize transients during system oscillations and keep the converter synchronized with the system. Note that the remote side frequency information could also be approximated using phasor domain methods [49]. In addition, estimated SV states of the system can also be used for other applications such as fast control of converters, harmonics filtering, etc.

## V. Protection Applications of DSE

When a fault occurs on power system components, protective relays need to immediately operate, to isolate the faulty components, minimize the power outage, and ensure the safety of human beings and the overall power system. Protective relays are evolving with improved reliability. However, statistically, the industry in the U.S. and abroad are still experiencing around 10 percent misoperation on average [50]. These are due to limited measurements, mis-coordination among relays, or faults that are hard to be quickly and reliably detected, such as high impedance faults, faults near neutrals, etc. In addition, most protective relays are based on fundamental frequency measurements, which limit their operational speed as well as applicability to DC systems. The proliferation of CBRs has also generated new challenges for legacy protection systems. Most relay settings are fixed; but with reduced system inertia, the settings need adjustment to distinguish stable and unstable swings. DSE based protection provides a new approach to deal with these challenges. DSE is a powerful tool in tracking power system transients (including electromechanical and electromagnetic transients) and therefore enables protective relays with improved operational speed, sensitivity, and reliability.

*A. DSE-based Protection using PMU Measurements*

*1) Out-of-step Protection using Direct Stability Assessment*

Quasi dynamic state estimation has been used to monitor the stability of generators and to detect instability when it occurs. This provides better system protection for stability. Specifically, generator instability is a serious problem for power systems; generators must be protected against this condition; out-of-step (OOS) relaying is used to detect and protect the generator when it spins into instability. Present out-of-step protective relays typically monitor the apparent impedance at the terminals of a generator. When instability occurs the impedance moves from the right-hand side of the impedance diagram to the left-hand side. Upon this detection, the relay will schedule to trip the generator when the generator swings another 120 to 150 degrees, which will minimize the transients in the breaker (transient recovery voltage) and allow the reliable opening of the breaker. Because of this timetable (i.e. detection of the instability and then a delay for favorable conditions to trip the generator), the period during which the generator experiences severe current flow due to the oscillation is long.

A new DSE based protection method has been introduced to detect the onset of instability much faster than the impedance-based approach described above [51]-[52]. The basic idea is quite simple: DSE is used to estimate the full dynamic state of the generator and the immediate network to which the generator is connected. The DSE provides the speed of the generator [49], the frequency, and RoCoF at each bus of the system. A simple calculation provides the center of oscillation (CoO). The CoO is characterized by the fact that the frequency at that point does not oscillate (it may be linearly changing with time) and the frequency at the nearest two buses oscillates with approximately 180 degrees out of phase. Knowledge of the generator speed and the location of the CoO enables the computation of the total energy of the generator,

$$E_{total} = E_p + E_k = -\int_{\delta_{coo}}^{\delta}(P_m - P_e(x))dx + \frac{1}{2}M\omega^2 \quad (2)$$

where $E_{total}$, $E_p$ and $E_k$ are the total energy, potential energy, and kinetic energy of the generator, respectively. $M = 2H/\omega_s$, $H$ is the per-unit inertia constant, $\delta$ is the generator position, $\omega$ is the rotor speed, and $\omega_s$ is the synchronous speed.

The stability of the generator is determined by the total energy using classical Lyapunov's theory. The theory states that when the total energy reaches the value of the peak potential energy (stability barrier) the generator becomes unstable. The dynamic state estimator provides the total energy of the generator as well as the potential energy function and asserts instability when the total energy reaches the stability barrier. It turns out that the assertion of generator instability occurs before the generator has swung away from the system and therefore can be tripped immediately, avoiding severe transients of an unstable generator. Next, we present a simplified example to illustrate the performance of this method.

The example test system is illustrated in Fig. 8. It consists of three generators, three transformers, six transmission lines, five loads, and 12 breakers. The parameters of the generators, transformers, lines, etc. are typical. We consider a fault on the 115 kV, a 38-mile-long line that is successfully cleared by the protection system of the line. The fault disturbs the system and oscillations are triggered.

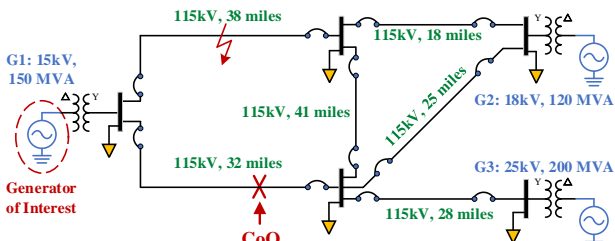

Fig. 8. Single-line diagram of the example test system

Fig. 9 illustrates the results of the DSE and computed trajectory of the impedance of the generator as "seen" at the terminal of the generator. The trajectory is superimposed on the characteristics of an OOS legacy protective relay to compare the DSE based protection results with the legacy protection. The apparent impedance moves to a value very close to the origin upon fault initiation at time t=1 sec. During the fault, as the generator and the system oscillate, the trajectory moves. The fault is cleared 0.25 sec after the occurrence of the fault (i.e. t=1.25 sec) by disconnecting the faulted transmission line. For this system, the critical clearing time is 0.2 sec. This means that the generator enters instability at time t=1.2 sec. The legacy relay will alarm the condition at time t=1.43 sec and will assert instability at t=1.51 sec while the generator rotor is at 216.2 degrees. At this angle, the generator cannot be tripped immediately (due to concerns over transient recovery voltage at the generator breaker and possibility of restrike) but need to wait until the phase angle goes to a smaller value. The DSE estimates the potential energy after the clearing of the fault and due to computational latencies, it asserts instability at t=1.29 sec when the generator rotor is at 118.4 degrees out of phase with the center of oscillation. At this phase angle, the generator can be tripped immediately, avoiding any additional stress on the generator.

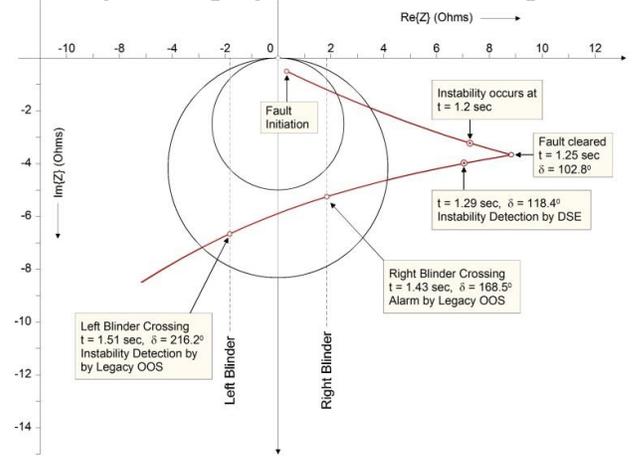

Fig. 9. Instability assertion time via DSE and comparison to legacy OOS Protection

*2) Adaptive OOS Relay Settings Approach*

Adaptive OOS settings based on quasi dynamic model parameters estimation preserves existing industrial practice and it is assisted by DSE. The OOS settings are mainly influenced by the direct axis transient reactance, quadrature axis speed voltage, and generator inertia. As seen by the relay, these parameters vary with time in the modern grid architecture with high penetration of power electronic interfaced technologies. The approach reported in [53] can be used to provide this real-time information for OOS relay settings recalculation. DSE is used to estimate dynamic model parameters which are used to recalculate the setting of OOS to adapt the relay sensitivity towards the generator instability with the system operating condition before the disturbance. This approach has been reported in [54]. In the paper, the settings of OOS are recalculated based on the extended equal area criterion method and demonstrated to be more effective OOS condition detection and protection.

### B. DSE-based Protection and Fault Location Using SV measurements

*1) DSE-based Protection Using SV Measurements*

Compared to the synchrophasor measurements from PMUs, synchronized SV measurements from state-of-the-art MUs capture system dynamics in time domain. DSE can be utilized to extract the system information embedded in SV measurements, to provide reliable detection of fault conditions that are not reflected by fault current levels, distortion of waveforms and characteristics of fault currents. In addition, the use of SV measurements provides detection of faulty conditions much faster than legacy protection functions (such as overcurrent protection, distance protection, current differential protection, etc.), which usually require collection of enough data to compute phasors, resulting in fault detection delays.

DSE based protection methods using SV measurements have been introduced in [7]. These methods were inspired from the widely adopted current differential protection. The latter examines whether the phasor domain KCLs of the protection zone is satisfied; many but not all internal faults will violate KCL allowing differential protection to detect these faults and trip the component. By contrast, DSE based protection examines whether all the physical laws of the specific protection zone are satisfied and an internal fault is detected with any violation of any physical law. Depending on the protection zone, physical laws may include KCLs, KVLs, motion laws, thermodynamic laws, etc. The primary improvements of the DSE based protection compared to the existing protection approaches (such as current differential protection) include: 1) speed: DSE based protection approach utilizes time-domain SV measurements instead of phasor/spectral domain synchrophasor measurements, which enables faster fault detection; 2) applicability to DC systems; 3) DSE based protection approach checks all the physical laws (instead of KCLs only), and therefore can detect internal faults with improved sensitivity and reliability. Note that current differential protection fails to detect some faults, for example, inter-turn faults in transformers, etc.

A systematic procedure for DSE-based protection is as follows: 1) building the dynamic model that encapsulates all the physical laws of the protection zone in the time domain; the model uses differential and algebraic equations (DAEs) that could include electromechanical, electromagnetic, and thermal transients; the model represents a high fidelity time-domain representation of the protection zone; 2) applying DSE for estimating dynamic states and checking the consistency between the available measurements and the dynamic model. Low consistency indicates that some of the physical laws are violated and therefore an internal fault is detected. The validity of the DSE based protection comes from the following key advantages of DSE: 1) accurately tracking the dynamics of the system; 2) systematically checking the consistency between the measurements and the dynamic model through the residuals, and 3) effectively filtering out measurement errors. DSE based protection schemes have been applied to transmission lines [55]-[56], microgrids [48], transformers [57], etc. With increased security and dependability compared to legacy methods, the DSE based protection can be utilized as the main protection of the component of interest (protection zone). Additionally, the DSE based protection is capable of detecting bad data through centralized substation protection (details in section V.B.3).

The implementation of an example DSE based protective relay with a two-terminal transmission line as the protection zone is shown in Fig. 10. Due to space limitations, the figure only demonstrates the relay on the left terminal of the line, with the inter-trip signal connected to the left side breaker (the relay on the right terminal is equivalent).

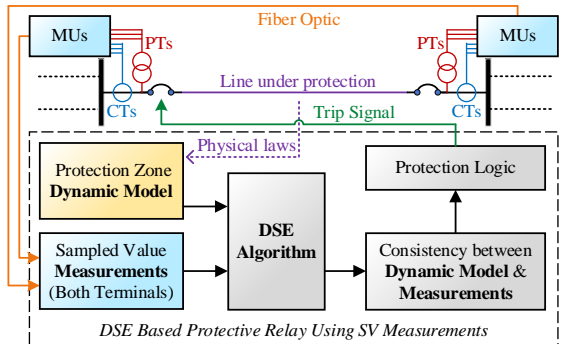

Fig. 10. Example DSE based protective relay using SV measurements at the left terminal, with a 2-terminal line as the protection zone

The two-terminal voltage and current SV measurements are obtained through PTs, CTs, and the MUs. The dynamic model of the protection zone is established by describing all the physical laws of the line under protection using DAEs. Afterward, DSE combines the dynamic model and the SV measurements, solves the dynamic states of the system, and computes the consistency between the dynamic model and the measurements (via the well-known chi-square test). Finally, the trip signal is issued to open the circuit breaker of the transmission line if an internal fault is detected. Using such an approach high impedance, or arcing faults are easily detected. A similar approach can be used to design smart auto-reclosure procedures [58].

*2) DSE-based Fault Location Using SV Measurements*

Fault Location is another important function when the fault occurs in a protection zone. Specifically, the exact fault location should be determined to minimize the time spent searching for the fault location by repair crews, yielding reduced power outage time and operating costs. With the development of fast-tripping techniques of protective relays, the time window of available measurements during faults for fault location is further shortened to the order of milliseconds. In addition, during this short time window, the available measurements usually experience severe transients. This leads to compromised accuracy of calculated phasors and therefore increases the fault location errors for legacy phasor domain-based fault location approaches.

To deal with these challenges, DSE based fault location approaches using SV measurements have been proposed in [59]-[61]. They were inspired by the widely adopted traditional phasor domain fault location methods. Traditional phasor domain methods build the relationship between the available phasor domain measurements and the location of the fault using algebraic equations, which are afterward solved to identify the location of the fault. DSE based fault location methods first describe the relationship between the available time-domain measurements and the location using an accurate time-domain dynamic model of the transmission line with fault. The time-domain model can also include the model of the arc [62]. The dynamic model is a set of DAEs, which typically include instantaneous voltages and currents through the transmission line as dynamic states of the system, and also the location of the fault as an extended state. Then, DSE is applied to systematically estimate the location of the fault. The primary advantages of DSE based fault location approach over traditional phasor domain approaches are 1) the DSE based methods make full use of information embedded in SV measurements and systematically filter out measurement errors; 2) time-domain algorithms typically possess faster convergence compared to phasor domain algorithms; 3) time-domain algorithms are not sensitive to complex harmonics distortions and are capable of accurate tracking severe dynamics during faults [62].

*3) Hidden Failure Detection of Protective Relays using DSE*

DSE-based protection schemes have also been applied at the substation level to achieve centralized substation protection [63], to detect compromised devices such as MUs and relays. The main idea is to utilize all the measurements in a substation, resulting in high data redundancy; this enables surgical detection of bad/missing data and validation of the source of anomalies (i.e. actual fault in the system, instrumentation faults, bad data injection cyber-attacks, etc.) via systematic hypothesis testing. Instrumentation faults (blown fuses, shorted CT, incorrect instrumental transformer ratios, etc.) known as hidden failures are a real problem in any protection system. DSE offers another advantage: upon detecting a hidden failure and bad data, the compromised data can be replaced in real-time by estimated values to ensure the resilient and reliable operation of the protection system.

## VI. Conclusions and Future Work

This paper has explored the usefulness and the advantages of DSE on many control and protection applications for modern power systems. It has been shown how DSE-based solutions comprehensively respond to challenges in the control and protection of modern power systems holistically. In addition, several gaps in the existing literature have also been identified and several new DSE-based control and protection methods have been proposed as possible solutions.

Future research on DSE applications on control and protection are categorized into the following key areas:

- **Role of AI/machine learning**: DSE enables the validation and calibration of system dynamic models, but there are still scenarios that good calibrations cannot be achieved due to the lack of high-quality large disturbances. As a result, the dynamic model may create deficiencies, yielding unreliable control, and protection. By mining the historical operational characteristics, advanced AI and machine learning tools may be able to compensate for these deficiencies. This further allows us to develop deep reinforcement learning algorithms that interact with the compensated DSE model outputs for system protection, control, and stability enhancement.

- **Enhancement of DSE and Observer**: Although the advantages and disadvantages of observer and DSE have been discussed, there is still a need to develop computationally efficient and robust alternatives in terms of handling sensor quality, model uncertainties, and nonlinearities. The practical implementation challenges for them need to be further investigated as well when designing state-feedback controls.

- **Management of distributed energy resources (DERs)**: In cases where fully decentralized DSE-based protection and control for all DERs do not cover all protection and control needs, as in cases of grid forming converters, multi-agent distributed DSE-based approaches need to be investigated, so as to share and transmit key information for frequency and voltage control, and protective actions.

- **Practical implementation**: Practical implementations should address issues of response time commensurate with the application. DSE applications need to be supported by adequate computing resources or distributed to achieve practical and acceptable performance in real-time. Many DSE applications of control and protection lack examples of their practical implementation in the field. Research topics include: How to ensure interoperability with currently used tools of control and protection? How to make them compatible with the current control room environment? What level of centralized or distributed computing is needed? What response times are desirable? What information should be communicated between remote sites and control rooms? What training should be developed and provided to power engineers? And what new standards are needed?

### Acknowledgment

The authors would like to thank Dr. Yingchen Zhang, Dr. Shahrokh Akhlaghi, and Dr. Anjan Bose for their valuable discussions.

### References

[1] J. Zhao, *et al.*, "Power system dynamic state estimation: motivations, definitions, methodologies, and future work," *IEEE Trans. Power Syst.*, vol. 34, no. 4, pp. 3188-3198, July 2019.

[2] J. Zhao, *et al.*, "Roles of dynamic state estimation in power system modeling, monitoring and operation," *IEEE Trans. Power Syst.*, 2020.

[3] V. Telukunta, J. Pradhan, A. Agrawal, M. Singh and S. G. Srivani, "Protection challenges under bulk penetration of renewable energy resources in power systems: A review", *CSEE Journal Power Energy Syst.*, vol. 3, no. 4, pp. 365-379, Dec. 2017.

[4] Protection System Misoperation Task Force, "Misoperations report", *NERC Planning Committee*, Atlanta, GA, USA, 2014.

[5] "Stability definitions and characterization of dynamic behavior in systems with high penetration of power electronic interfaced technologies," *IEEE Technical Report, IEEE Power and Energy Society, Power System Dynamic Performance Committee*, 2020.

[6] IEC Std 61850, "Communication Networks and Systems in Substations", 2003.

[7] A. P. Meliopoulos et al., "Dynamic state estimation-based protection: status and promise," *IEEE Trans. Power Delivery*, vol. 32, no. 1, pp. 320-330, Feb. 2017.

[8] "1,200 MW Fault Induced Solar Photovoltaic Resource Interruption Disturbance Report: Southern California 8/16/2016 Event", *NERC*, Atlanta, GA, June 2017.

[9] Z. Huang, H. Krishnaswami, G. Yuan, and R. Huang, "Ubiquitous power electronics in future power systems", *IEEE Electrification Magzine*, September 2020.

[10] V. Venkatasubramanian, H. Schattler, and J. Zaborszky, "Fast time-varying phasor analysis in the balanced three-phase large electric power system," *IEEE Trans. on Automatic Control*, vol. 40, no. 11, pp. 1975-1982, Nov. 1995.

[11] S. R. Sanders, J. M. Noworolski, X. Z. Liu, G. C. Verghese, "Generalized averaging method for power conversion circuits," *IEEE Trans. on Power Elec.*, vol. 6, no. 3, pp. 251-259, April 1991.

[12] H. K. Khalil, *Nonlinear Systems*, 3rd ed., Prentice-Hall, 2002.

[13] M. Korda and I. Mezić, "Linear predictors for nonlinear dynamical systems: Koopman operator meets model predictive control," *Automatica*, vol. 93, pp. 149–160, Jul. 2018.

[14] M. Nicolai, L. Lorenz-Meyer, A. Bobtsov, R. Ortega, N. Nikolaev, and J. Chiffer, "PMU-based decentralized mixed algebraic and dynamic state observation in multi-machine power systems", *IET Gener. Transm. Distrib.*, accepted.

[15] J. B. Zhao, L. Mili, "A robust generalized-maximum likelihood unscented Kalman filter for power system dynamic state estimation," *IEEE Journal of Selected Topics in Signal Processing*, vol. 12, no. 4, pp. 578-592, 2018.

[16] J. Qi, A.F. Taha, J. Wang, "Comparing Kalman filters and observers for power system dynamic state estimation with model uncertainty and malicious cyber-attacks," *IEEE Access*, vol. 6, pp. 77155-77168, 2018.

[17] S. Nugroho, A.F. Taha, J. Qi, "Robust dynamic state estimation of synchronous machines with asymptotic state estimation error performance guarantees," *IEEE Trans. Power Syst.*, vol. 35, no. 3, pp. 1923-1935, 2020.

[18] S. Stefani, H. Savant, *Design of Feedback Systems*, Oxford University Press, New York, 2002

[19] A.F. Taha, M. Bazrafshan, S. A. Nugroho, N. Gatsis, J. Qi, "Robust control for renewable-integrated power networks considering input bound constraints and worst-case uncertainty measure," *IEEE Trans. Contr. Network Syst.*, 2019

[20] A. K. Singh, R. Singh, B.C. Pal, "Stability analysis of networked control in smart grids," *IEEE Trans. Smart Grid*, vol. 6, no. 1, pp. 381-390, Jan. 2015.

[21] A.M. Ersdal, L. Imsland, K. Uhlen, "Model predictive load frequency control," *IEEE Trans. Power Syst.*, vol. 31, no. 1, pp. 777-785, Jan. 2016.

[22] A. Paul, I. Kamwa and G. Jóos, "Centralized Dynamic State Estimation Using a Federation of Extended Kalman Filters With Intermittent PMU Data From Generator Terminals," *IEEE Trans. Power Syst.*, vol. 33, no. 6, pp. 6109-6119, Nov. 2018.

[23] E. Ghahremani, I. Kamwa, "Local and wide-area PMU-based decentralized DSE in multi-machine power systems," *IEEE Trans. Power Syst.*, vol. 31, no. 1, pp. 547-562, 2016.

[24] W. Yao, L. Jiang, J. Wen, Q. H. Wu, S. Cheng, "Wide-area damping controller of FACTS devices for inter-area oscillations considering communication time delays," *IEEE Trans. Power Syst.*, vol. 29, no. 1, pp. 318-329, Jan. 2014.

[25] A. Paul, I. Kamwa, G. Joos, "PMU signals responses-based RAS for instability mitigation through on-the fly identification and shedding of the run-away generators," *IEEE Trans. Power Syst.*, vol. 35, no. 3, pp. 1707-1717, May 2020.

[26] A. K. Singh, B. C. Pal, "Decentralized control of oscillatory dynamics in power systems using an extended LQR," *IEEE Trans. Power Syst.*, vol. 31, no. 3, pp. 1715-1728, May 2016.

[27] A. K. Singh, B. C. Pal, "Decentralized nonlinear control for power systems using normal forms and detailed models," *IEEE Trans. Power Syst.*, vol. 33, no. 2, pp. 1160-1172, Mar. 2018.

[28] S. Liu, X. Li, D. Chen, "Wide-area-signals-based nonlinear excitation control in multi-machine power systems", *IEEJ Trans. Elec. Electron. Eng.*, vol. 14, pp. 366-375, 2019.

[29] H. Liu, J. Su, J. Qi, N. Wang, C. Li, "Decentralized voltage and power control of multi-machine power systems with global asymptotic stability," *IEEE Access*, vol. 7, pp. 14273-14282, 2019.

[30] A. S. Mir, S. Bhasin, N. Senroy, "Decentralized nonlinear adaptive optimal control scheme for enhancement of power system stability," *IEEE Trans. Power. Syst*, vol. 35, no. 2, pp. 1400-1410, 2020.

[31] A. K. Singh, B. C. Pal, "Rate of change of frequency estimation for power systems using interpolated DFT and Kalman filter," *IEEE Trans. Power Syst.*, vol. 34, no. 4, pp. 2509-2517, 2019.

[32] H. Bevrani, *Robust Power System Frequency Control*. New York, NY, USA: Springer, 2009.

[33] P. Bhui, N. Senroy, A. K. Singh and B. C. Pal, "Estimation of inherent governor dead-band and regulation using unscented Kalman filter", *IEEE Trans. Power Syst.*, vol. 33, no. 4, pp. 3546-3558, 2018.

[34] A. S. Mir and N. Senroy, "Self-tuning neural predictive control scheme for ultra-battery to emulate a virtual synchronous machine in autonomous power systems," *IEEE Trans. Neural Netw. Learn. Syst.*, vol. 31, no. 1, pp. 136-147, Jan. 2020.

[35] K. Emami, T. Fernando, H. H. Iu, B. D. Nener, K. P. Wong, "Application of unscented transform in frequency control of a complex power system using noisy PMU data" *IEEE Trans. Ind. Informat.*, vol. 12, no. 2, pp. 853-863, 2016.

[36] H. Trinh, T. Fernando, H. H. C. Iu, K. P. Wong, "Quasi-decentralized functional observers for the LFC of interconnected power systems", *IEEE Trans. Power Syst.*, vol. 28, no. 3, pp. 3513-3514, 2013.

[37] M. Glavic, T. Van Cutsem, "Wide-area detection on voltage instability from synchronized phasor measurements. Part I: Principle," *IEEE Trans. Power Syst.*, vol. 24, no. 3, pp. 1408-1416, 2009.

[38] B. A. Alcaide-Moreno, C. FR. Fuerte-Esquivel, M. Glavic, T. Van Cutsem, "Electric power network state tracking from multirate measurements," *IEEE Trans. Instrumentation and Measurement*, vol. 67, no. 1, pp. 33-44, 2018.

[39] G. K. Morison, B. Gao, P. Kundur, "Voltage stability analysis using static and dynamic approaches," *IEEE Trans. Power Syst.*, vol. 8, no. 3, pp. 1159-1171, 1993.

[40] M. Glavic, T. Van Cutsem, "Some reflections on model predictive control of transmission voltages," *38th North American Power Symposium, Carbondale, IL, USA*, 2006.

[41] B. Otomega, M. Glavic, T. Van Cutsem, "A two-level emergency control scheme against power system voltage instability," *Control Engineering Practice*, vol. 30, pp. 93-104, 2014.

[42] S. Yu, T. Fernando, K. Emami and H. H.-C. Iu, "Dynamic state estimation-based control strategy for DFIG wind turbine connected to complex power systems," *IEEE Trans. Power Syst.*, vol. 32, no. 2, pp. 1272-1281, 2016.

[43] A. S. Mir and N. Senroy, "DFIG damping controller design using robust CKF based adaptive dynamic programming," *IEEE Trans. Sustain. Energy*, vol. 11, no. 2, pp. 839-850, Apr. 2020.

[44] S. Yu, G. Zhang, T. Fernando and H. H.-C. Iu, "A DSE-based SMC method of sensorless DFIG wind turbines connected to power grids for energy extraction and power quality enhancement," *IEEE Access*, no. 6, pp. 76596-76605, 2018.

[45] S. Yu, L. Zhang, H. H. Lu, T. Fernando, K. P. Wong, "A DSE-based power system frequency restoration strategy for PV-integrated power systems considering solar irradiance variations," *IEEE Trans. Industrial Informatics*, vol. 13, no. 5, pp. 2511-2518, 2017.

[46] "Wind Energy Systems Subsynchronous Oscillations: Events and Modeling," IEEE PES Wind SSO Taskforce, Technical Report, PES-TR80, 2020.

[47] A. P. Meliopoulos, V. Vittal, M. Saeedifard, and R. Data, "Stability, protection and control of systems with high penetration of converter interfaced generation", PSERC Publication 16-03, March 2016.

[48] Y. Liu, A. P. Meliopoulos, L. Sun, and S. Choi, "Protection and control of microgrids using dynamic state estimation", *Protection Control Modern Power Syst.*, vol. 3, no. 31, pp. 1-13, Oct. 2018.

[49] R. Grondin A. Heniche, *et al.,* "Loss of synchronism detection a strategic function for power system protection", *Proc. 2006 CIGRE Session* 41 B5-205.

[50] North American Electric Reliability Corporation, "State of Reliability 2016", May 2016.

[51] E. Farantatos, R. Huang, G. J. Cokkinides, A. P. Meliopoulos, "A predictive generator out-of-step protection and transient stability monitoring scheme enabled by a distributed dynamic state estimator," *IEEE Trans. Power Del.*, vol. 31, no. 4, pp. 1826-1835, Aug. 2016.

[52] Y. Cui, R. G Kavasseri, and S. M Brahma, "Dynamic state estimation assisted out-of-step detection for generators using angular difference," *IEEE Trans. Power Del.*, vol. 32, no. 3, pp. 1441–1449, Jun. 2017.

[53] M. A. M. Ariff, B. C. Pal and A. K. Singh, "Estimating dynamic model parameters for adaptive protection and control in power system," *IEEE Trans. Power Syst.*, vol. 30, no. 2, pp. 829-839, March 2015.

[54] M. A. M. Ariff and B. C. Pal, "Adaptive protection and control in the power system for wide-area blackout prevention," *IEEE Trans. Power Del.*, vol. 31, no. 4, pp. 1815-1825, Aug. 2016.

[55] Y. Liu, A. P. Meliopoulos, R. Fan, L. Sun, Z. Tan, "Dynamic state estimation based protection on series compensated transmission lines", *IEEE Trans. Power Del.*, vol. 32, no. 5, pp 2199-2209, Oct. 2017.

[56] Y. Liu, A. P. Meliopoulos, L. Sun and R. Fan, "Dynamic state estimation based protection on mutually coupled transmission lines," *CSEE Journal Power Energy Syst.*, vol. 2, no. 4, pp. 6-14, Dec. 2016.

[57] R. Fan, Y. Liu, A. P. Meliopoulos, L. Sun, Z. Tan and R. Huang, "Comparison of transformer legacy protective functions and a dynamic state estimation-based approach", *Electric Power Syst. Research*, 2020.

[58] V. Terzija, G. Preston, V. Stanojević, N. I. Elkalashy, and M. Popov, "Synchronized measurements-based algorithm for short transmission line fault analysis", *IEEE Trans Smart Grid*, vol. 6, no. 17, pp. 2639-2648, Nov. 2015.

[59] Y. Liu, A. P. Meliopoulos, Z. Tan, L. Sun and R. Fan, "Dynamic state estimation-based fault locating on transmission lines", *IET Gener. Transm. Distrib.*, vol. 11, no. 17, pp. 4184-4192, Nov. 2017.

[60] R. Fan, Y. Liu, R. Huang, R. Diao and S. Wang, "Precise fault location on transmission lines using ensemble Kalman filter", *IEEE Trans. Power Del.*, vol. 33, no. 6, pp. 3252-3255, Dec. 2018.

[61] B. Wang, Y. Liu, D. Lu, K. Yue and R. Fan, "Transmission line fault location in MMC-HVDC grids based on dynamic state estimation and gradient descent", *IEEE Trans. Power Del.*, 2020.

[62] M. B. Djuric, Z. M. Radojevic, and V. V. Terzija. "Time domain solution of fault distance estimation and arcing faults detection on overhead lines," *IEEE Trans. Power Del.*, vol 14, no. 1, pp. 60-67, Feb. 1999.

[63] H. F. Albinali and A. P. Meliopoulos, "Resilient protection system through centralized substation protection," *IEEE Trans. Power Del.*, vol. 33, no. 3, pp. 1418-1427, June 2018.